%% file: main.tex
\documentclass[12pt]{article}

\pdfoutput=1

\usepackage[margin=1in]{geometry}

\usepackage{amsmath, amsthm, amssymb, mathtools}
\usepackage{thmtools}
\usepackage{thm-restate}
\usepackage[utf8]{inputenc}
\usepackage[shortlabels]{enumitem}
\usepackage[
  bookmarks=true,
  bookmarksnumbered=true,
  bookmarksopen=true,
  pdfborder={0 0 0},
  breaklinks=true,
  colorlinks=true,
  linkcolor=black,
  citecolor=black,
  filecolor=black,
  urlcolor=black,
]{hyperref}
\usepackage{xurl}
\usepackage[table,x11names]{xcolor}
\usepackage{tikz}
\usetikzlibrary{decorations.pathmorphing}

\usetikzlibrary{arrows.meta, positioning, fit, backgrounds}
\usepackage{nicefrac}
\usepackage{setspace}
\newlength{\boundfracwd}
\newcommand{\bfrac}[2]{\makebox[\boundfracwd][c]{$\frac{#1}{#2}$}}

\usepackage[capitalise]{cleveref}

\theoremstyle{plain}
\newtheorem{theorem}{Theorem}
\numberwithin{theorem}{section}

\newtheorem*{open-question}{Open Question}

\theoremstyle{definition}

\newtheorem{example}[theorem]{Example}

\newcommand{\N}{\mathbb{N}}

\usepackage{algpseudocode}
\usepackage{algorithm}
\usepackage{footnote}
\makesavenoteenv{algorithm}

\newcommand{\problem}{\bigl(N,G,(\succ_i)_{i\in N}\bigr)}

\renewcommand{\epsilon}{\varepsilon}

\usepackage{fvextra}
\DefineVerbatimEnvironment{prompt}{Verbatim}
  {fontsize=\small, xleftmargin=2.5em, xrightmargin=2.5em,
   breaklines=true, breakanywhere=true,
   breaksymbolleft={}, breaksymbolright={}}

\usepackage[backend=biber,hyperref,doi,style=authoryear,url=false,
sorting=nyt,natbib=true,giveninits=true,uniquename=init]{biblatex}
\AtEveryBibitem{
\clearfield{note}
\clearfield{isbn}
\clearfield{issn}
\clearlist{publisher}
\clearfield{series}
\clearfield{eventtitle}
\clearlist{location}
\clearfield{language}
}

\renewbibmacro*{textcite}{%
  \iffieldundef{shorthand}
    {\ifnameundef{labelname}
       {\usebibmacro{cite:label}%
        \setunit{%
          \global\booltrue{cbx:parens}%
          \printdelim{nonameyeardelim}\bibopenparen}%
        \ifnumequal{\value{citecount}}{1}
          {\usebibmacro{prenote}}
          {}%
        \usebibmacro{cite:labeldate+extradate}}
       {\printnames{labelname}%
        \setunit{%
          \global\booltrue{cbx:parens}%
          \printdelim{nameyeardelim}\bibopenparen}%
        \ifnumequal{\value{citecount}}{1}
          {\usebibmacro{prenote}}
          {}%
        \usebibmacro{citeyear}}}
    {\usebibmacro{cite:shorthand}}}

\title{Stable Menus of Public Goods: AI-Enabled Progress\thanks{Accepted to the \href{https://sites.google.com/view/aieconcs26/}{EC'26 Workshop on AI-Driven Research in EconCS}.}}

\author{Sara Fish\thanks{School of Engineering and Applied Sciences, Harvard University. Fish was supported by an NSF Graduate Research Fellowship and a Kempner Institute Graduate Fellowship. | \emph{E-mail}: \mbox{\href{mailto:sfish@g.harvard.edu}{sfish@g.harvard.edu}.} Thanks to Yannai Gonczarowski and Pras Ramakrishnan for helpful comments.}}

\date{June 15, 2026}

\begin{document}

\maketitle

\begin{abstract}

Using an open problem from the EC 2025 paper \emph{``Stable Menus of Public Goods''} as a testbed, we conduct experiments to understand the effectiveness of different AI-for-EconCS research workflows. Specifically, we study three questions: Does providing human intuition in the prompt help? Does automated multi-turn interaction help? And, does an LLM outperform a first-year PhD student? Regarding the first two questions, we provide evidence for the following workflow suggestions: (1) prompting with human intuition can encourage the LLM to have better ``taste'', (2) multi-turn workflows help when the pipeline encourages ``ambitious'' steps. Regarding the third question, using an unpublished manuscript written by the paper's senior authors prior to collaborating with the first-year PhD student, we compare the effectiveness of the LLM with that of the first-year PhD student, and find that the LLM is slightly less effective. 

\end{abstract}

\section{Introduction}

Inspired by recent AI-driven leaps in combinatorics \citep{openai_openai_2026,bloom_sum-product_2026}, we revisit an open problem from the EC 2025 paper \emph{``Stable Menus of Public Goods''} (\citeauthor{fish_2025_stable}, \citeyear{fish_2025_stable}). \citet{fish_2025_stable} consider a novel public goods model that exhibits a rich combinatorial structure. For the question of the existence of stable menus in this model, they prove non-matching lower and upper bounds, and leave closing this gap as an open problem. 

In this paper, we use this open problem from \citet{fish_2025_stable} as a testbed for evaluating the effectiveness of different AI-for-EconCS research workflows. Specifically, we conduct experiments to shed light on three research questions:

  \begin{description} 
    \item[\textbf{(RQ1)}] Does providing human intuition in the prompt help?
    \item[\textbf{(RQ2)}] Does automated multi-turn interaction help?
    \item[\textbf{(RQ3)}] Does an LLM outperform a first-year PhD student? 
  \end{description}

To study \textbf{(RQ1)}, in \cref{sec:human_intuition}, we conduct experiments in which an LLM (GPT-5.5 Pro with Extended Thinking) is asked to improve on the lower or upper bounds from \cite{fish_2025_stable} in a single query. We compare two prompt types: \textsc{WithContext}, in which human intuition about what directions might be most promising is provided, and \textsc{NoContext}, which includes no such information. We observe that providing human intuition \emph{per se} does not appear to help much, but that (for upper bounds) it appears to encourage the LLM to be more ``ambitious'' / have better ``taste'', which leads to improved results.

To study \textbf{(RQ2)}, in \cref{sec:autoresearcher}, we construct a multi-turn pipeline in which a ``Supervisor'' agent (Claude 4.7 Opus) repeatedly encourages a ``Researcher'' agent (GPT-5.5 Pro with ``xhigh'' effort) to improve on its own lower or upper bounds. We test three rollouts of this pipeline, each consisting of ten turns. One rollout, seeking lower bounds, successfully arrived at results better than those from the single-shot pipeline from \cref{sec:human_intuition}. However, the other two rollouts got ``stuck'', whereby the Supervisor often made low-quality, incremental suggestions. These findings suggest that Supervisor-type agents may be more effective when nudged to be more ambitious and high-level in their feedback. 

To study \textbf{(RQ3)}, in \cref{sec:first-year}, we conduct experiments in which the LLM (GPT-5.5 Pro with Extended Thinking) is instead asked to improve on GH'20, an unpublished manuscript documenting the senior authors G and H's progress on the problem prior to F joining the collaboration. By comparing the LLM's improvements on GH'20 with \cite{fish_2025_stable}, we can compare the relative effectiveness of the LLM with that of the author F (myself as a first-year PhD student). Overall, we find that the LLM is slightly less effective than the first-year PhD student; however, the PhD student's advantage appears to be fragile.

Finally, regarding the actual problem at hand, the autoresearcher pipeline from \cref{sec:autoresearcher} improves \citet{fish_2025_stable}'s lower bound (necessary condition, higher is better) result of $\frac{u-1}{t-1} \gtrapprox 23/11 \approx 2.09$ to $\frac{u-1}{t-1} \gtrapprox 8/3 \approx 2.67$, and the single-shot \textsc{WithContext} prompt from \cref{sec:human_intuition} improves \citet{fish_2025_stable}'s upper bound (sufficient condition, lower is better) result of $\frac{u-1}{t-1} \ge g-2$ to $\frac{u-1}{t-1} \gtrapprox \frac{g}{2} + \log_2(g+1)$. In my opinion, both of these directions constitute substantial improvements that are interesting in their own right: the LLM's lower bound additionally resolves a separate (minor) open question from \citet{fish_2025_stable}, and the LLM's upper bound uses novel techniques that go beyond those used in \citet{fish_2025_stable}.

\paragraph{Related Work.} This work relates to the growing literature on AI for Science \citep{wang_scientific_2023}. Works including \citet{georgiev_mathematical_2025,bloom_sum-product_2026,nagda_reinforced_2026,openai_openai_2026} demonstrate the promise of using LLMs to advance combinatorics research (see also \citealt{wagner2021constructions,charton_patternboost_2024}, which use non-LLM ML methods). The multi-turn ``autoresearcher'' pipeline (\cref{sec:autoresearcher}) follows directly in the footsteps of more elaborate systems such as \citet{novikov2025alphaevolve,breen_ax-prover_2026,feng_towards_2026,tsoukalas_advancing_2026,zheng2026aicomathematicianacceleratingmathematicians}. This paper's contribution is to document and ablate certain AI-for-Science workflow choices, using a specific open problem from \cite{fish_2025_stable} as a benchmark.

\section{Background}\label{background}

In this section we briefly describe the open problem the AI is asked to make progress on. For a more fleshed-out exposition, see \citet{fish_2025_stable}. For a less fleshed-out exposition, it suffices to look at \cref{fig:bounds}. Very briefly (for problem parameters $g,t,u$), the problem is solved for $g \le 6$; for $g \ge 7$, lower bound (necessary condition) improvements involve providing constructions that beat $\frac{u-1}{t-1} \gtrapprox \frac{23}{11}$ (i.e., value exceeding $\frac{23}{11} \approx 2.1$), and upper bound (sufficient condition) improvements involve providing a proof that beats $\frac{u-1}{t-1} \ge g-2 $ (i.e., value lower than $g-2$).

\subsection{Problem Statement}

 A \textbf{menu selection problem} is a triplet $\problem$ where $N = \{1,\ldots,n\}$ is a finite set of agents, $G = \{1,\ldots,g\}$ is a finite set of (public) goods, and agent $i$'s (possibly incomplete) preference relation $\succ_i$ is a strict partial order over $G$. A \textbf{menu} $O$ is a set of goods. Given a menu $O$, the \textbf{agent assignment} $a_O: N \to O \cup \{ \perp \}$ maps each agent to their favorite good in $O$ (or an outside option $\perp$ if the agent's preference list does not contain any goods in $O$). The intuition is that a central planner selects a menu of goods $O$ to offer, and then each (unit-demand) agent uses the good from $O$ that they prefer most.

Given parameters $t, u \in \N$, a menu $O$ is \textbf{$t$-feasible} if every offered good is the assigned good of at least $t$ agents, that is, $\bigl|a_O^{-1}(j)\bigr| \ge t$ for every $j \in O$. A menu $O$ is \textbf{$u$-uncontestable} if no unoffered good is preferred to all of $O$ by $u$ or more agents---that is, $\bigl|a_{O \cup \{j\}}^{-1}(j)\bigr| < u$ for every $j \in G \setminus O$. The intuition behind $t$-feasibility is that every good offered in the menu $O$ should ``justify its existence'' by being used by at least $t$ agents; the intuition behind $u$-uncontestability is that the menu $O$ should not fail to include a popular good that at least $u$ agents would have preferred over all other goods in $O$.

A menu is \textbf{$(t,u)$-stable} if it is both $t$-feasible and $u$-uncontestable. One of the central questions of \citet{fish_2025_stable} is: for which triples $(g, t, u)$ does every menu selection problem on $g$ public goods admit a $(t,u)$-stable menu?

\begin{example}
Consider a menu selection problem with $n = 9$ agents and $g = 3$ goods (denote the set of goods $G := \{1,2,3\}$), where three agents have preferences $1 \succ 2 \succ 3$, three agents have preferences $2 \succ 3 \succ 1$, and three agents have preferences $3 \succ 1 \succ 2$.

Fix $t = 4$ and $u = 7$. Then $O := \{ 1 \}$ is $(t,u)$-stable, because it is $t$-feasible (all $9 \ge t$ agents use good 1) and $u$-uncontestable (only $6 < u$ agents prefer good 3 over good 1).

However, if we instead set $u = 6$, then no $(t,u)$-stable menus exist. The menu $O := \{1 \}$ is $u$-uncontestable since $6 \ge u$ agents prefer good 3 over good 1. (By symmetry, $\{ 2 \}$ and $\{ 3 \}$ are similarly $u$-uncontestable.) The menu $O := \{1,2 \}$ is $t$-infeasible since only $ 3< t$ agents use good 2. (By symmetry, $\{ 2,3\}$ and $\{ 1,3 \}$ are similarly $t$-infeasible.) Finally, by monotonicity, $O := \{1,2,3 \}$ is $t$-infeasible and $O := \emptyset$ is $u$-uncontestable.

\end{example}

More generally, $(t,u)$-stability as a condition becomes stronger as $t$ increases and as $u$ decreases (and vice versa). Thus, for a given $g$ and $t$, the question is to find the ``cutoff'' $u_g(t)$ so that for $u \ge u_g(t)$, every menu selection problem on $g$ goods has a $(t,u)$-stable menu, and for $u < u_g(t)$, there exists a menu selection problem on $g$ goods with no $(t,u)$-stable menu.

\subsection{Known Results}

The existence bounds obtained in \citet{fish_2025_stable} are as follows (see \Cref{fig:bounds} for a visualization):
\begin{itemize}
  \item (Proposition 3.1 and 1.2) For $g = 2$ and $t,u \in \mathbb{N}$, every menu selection problem has a $(t,u)$-stable menu if and only if $\frac{u-1}{t-1} \ge 1$.
  \item (Proposition 1.1 and 1.2) For every $g \in \{3,4,5,6\}$ and $t,u \in \mathbb{N}$, every menu selection problem has a $(t,u)$-stable menu if and only if $\frac{u-1}{t-1} \ge 2$.
  \item (Theorem 1.4) For every $g \ge 7$, there exists a menu selection problem with no $(t,u)$-stable menu if $u \le 23 \lfloor \frac{t-1}{11} \rfloor$ (roughly $\frac{u-1}{t-1} < \frac{23}{11} \approx 2.1$). That is, $u \ge 23 \lfloor \frac{t-1}{11} \rfloor + 1$ is a necessary condition for the existence of $(t,u)$-stable menus. 
  \item (Theorem 1.5) For every $g \ge 7$, every menu selection problem \emph{on complete preferences} has a $(t,u)$-stable menu if $\frac{u-1}{t-1} \ge g-2$.
  \item (Proposition E.4) For every $g \ge 5$, every menu selection problem---including those with \emph{incomplete} preferences---has a $(t,u)$-stable menu if $u > (g-1-\frac16)(t-1)$ (equivalently, $u \ge \lfloor (g - \frac76)(t-1) \rfloor + 1$).
\end{itemize}

That is, for $g \in \{2,3,4,5,6 \}$, the paper provides a complete characterization, and for $g \ge 7$, the paper provides (non-matching) necessary and sufficient conditions.

\input{baselines_figure.tex}

\paragraph{Notation.} Many results, both in \citet{fish_2025_stable} and proven by the LLM in this paper, arrive at bounds of the form $u-1 \ge k_1 \lfloor \frac{t-1}{k_2} \rfloor $, for $k_1,k_2 \in \mathbb{N}$. For the rest of this paper, we use the notation $\frac{u-1}{t-1} \gtrapprox \frac{k_1}{k_2}$ (or $\gtrapprox {k_1}/{k_2}$ for short) to refer to such inequalities. (When $k_2$ divides $t-1$, they are equivalent when $\gtrapprox$ is replaced by $\ge$, but otherwise, $u-1 \ge k_1 \lfloor \frac{t-1}{k_2} \rfloor $ is slightly weaker than $\frac{u-1}{t-1} \ge \frac{k_1}{k_2}$.)

\section{Does providing human intuition in the prompt help?}\label{sec:human_intuition}

We first investigate whether human intuition can be leveraged to elicit stronger results from the LLM. To do so, we compare two kinds of prompts. The first prompt type, \textsc{NoContext}, simply provides a link to the paper and asks the LLM to improve on the paper's results. The second prompt type, \textsc{WithContext}, additionally includes a description of the paper's results and some hints about potentially promising directions.\footnote{The \textsc{WithContext} hints were written based on deep familiarity with \citet{fish_2025_stable}, but had not been seriously pursued in advance. That is, they were written purely based on where the paper ``left off'', without prior knowledge of what improvements were attainable.} We write separate (but similar) prompts for the lower and upper bound cases, leading to four different prompts. (See \cref{prompts_single_turn} for the full prompts.)

For example, \textsc{NoContextUpper} reads (in its entirety):

\noindent\begin{minipage}{\linewidth}
\begin{prompt}
https://arxiv.org/abs/2402.11370

I would like you to work on deriving improvements for the upper bounds, that is, the sufficient condition bounds. Do not work on lower bound constructions, because another agent has that covered. Keep working until you have an actual upper bound proof better than the bounds in the paper. No need to report back with partial progress not yet constituting a proof. All of the upper bounds in the paper in fact are very loose, so progress is tractable.
\end{prompt}
\end{minipage}

Whereas \textsc{WithContextUpper} includes context such as:

\begin{prompt}
[...] A good place to start would be to look at the upper bound proofs in the paper. They include the aforementioned (u-1)/(t-1) = g-2 bound, but also the simple u >= (g-1)(t-1) argument in Appendix E, as well as the more refined u >= (g-1-eps)(t-1), also in Appendix E. These arguments arrive at sufficient conditions for the existence of (t,u)-stable menus for certain g by uncovering some sort of mathematical structure about the problem. The (g-2) bound uncovers one kind of structure (this "no gaps" idea), and the (g-1-eps) bound uncovers a different kind of structure (that, when small (t,u)-stable menus fail to exist, for u large, this implies that the menu selection problem has a specific cyclic structure). One promising direction could be to push either of these observations, or both simultaneously. Or, you are welcome to pursue other techniques. [...]
\end{prompt}

Some further remarks on the prompts:

\begin{itemize}
  \item The instruction ``Do not work on lower bound constructions'' was added because in early testing, when instructed to prove upper bounds, the LLM would often ``give up'' and work on lower bound constructions instead. 
  \item  The instruction ``All of the upper bounds in the paper in fact are very loose, so progress is tractable'' was similarly added to ``encourage'' the LLM. 
  \item The prompt includes the paper URL, rather than the entire paper \texttt{.tex} or PDF, to allow the LLM to read the paper with the method it most ``prefers''. (In all cases, the LLM has web browsing enabled.) In practice, the LLM appears to use a mix of PDF and HTML. 
\end{itemize}

Using GPT-5.5 Pro with Extended Thinking via the ChatGPT web interface, we collect five samples each of \textsc{NoContextLower}, \textsc{NoContextUpper}, \textsc{WithContextLower}, and \textsc{WithContextUpper}. Due to monthly Pro query limits, the samples were collected across two accounts. Browsing was enabled. All personalization features were left blank or set to default values, and memory was disabled.  

\paragraph{Lower bounds.} The lower bound results are displayed in \cref{tab:llm_lower_bounds}. A link to the full LLM outputs is provided in \cref{app:code_data_release}. All proofs (constructions) were checked by a human. 

Across all ten trials, the LLM improves on the $g \ge 7$ lower bound from \citet{fish_2025_stable}. Providing human intuition in the prompt (\textsc{WithContext}) reduces the time taken by the LLM by 18m11s on average ($p = 0.023 < 0.05$, two-sided Welch's $t$-test). However, the \textsc{NoContext} and \textsc{WithContext} prompts produce equally strong bounds on $\frac{u-1}{t-1}$ (in both cases, improving to $\gtrapprox 2.5$ four times and $\gtrapprox 7/3 \approx 2.3$ one time). 

All ten of the LLM's proofs follow a similar blueprint. The LLM produces a construction on $g=9$ goods, with a similar cyclic structure to the construction on $g=7$ goods from \citet{fish_2025_stable}. The LLM finds this construction using a mix of LP/MILP solvers and heuristics (e.g.~imposing cyclic structure). The proof that no $(t,u)$-stable menu exists also follows the methods of \citet{fish_2025_stable}: in their $g = 7$ construction, they identify a \emph{(3,4)-gap} (no $O \subseteq G$ with $|O|=3$ is $u$-uncontestable, and no $O \subseteq G$ with $|O| = 4$ is $t$-feasible), and in the LLM's $g = 9$ constructions, it identifies a $(4,5)$-gap (analogously, no $O \subseteq G$ with $|O|=4$ is $u$-uncontestable, and no $O \subseteq G$ with $|O|=5$ is $t$-feasible). 

\input{lower_bounds.tex}

\paragraph{Upper bounds.} The upper bound results are displayed in \cref{tab:llm_upper_bounds}. A link to the full LLM outputs is provided in \cref{app:code_data_release}. All proofs were checked by a human.

\input{upper_bounds.tex}

Across all ten trials, the LLM \emph{technically} improves on the upper bounds from \citet{fish_2025_stable}, however, in eight out of these ten cases the improvements are near-trivial. The four $\gtrapprox g-2 + \frac{5}{g+1}$ ``improvements'' are completely trivial: Proposition E.4 of \citet{fish_2025_stable} bounds $\frac{5}{g+1}$ by $\frac{5}{6}$ (using $g \ge 5$) to make the theorem statement cleaner, and the LLM's improvement is simply to restore the $\frac{5}{g+1}$ term in the bound. The four $\ge g-2$ improvements follow from observing that Lemma 5.7 from \citet{fish_2025_stable}, which was thought to only hold for complete preferences, can be extended to the incomplete preferences case (with a relatively simple argument, that admittedly was overlooked by \citealt{fish_2025_stable}).

Both of the two nontrivial improvements are achieved by \textsc{WithContext} prompts.\footnote{Trial 3 of \textsc{NoContextUpper} additionally ``proves'' an upper bound of $\ge g-3$, but it relies on an assumption that makes the problem substantially easier, so we do not consider this progress.} Trial 3 achieves a slight improvement (replacing $\frac{5}{g+1}$ with $\frac{4}{g}$) by tightening the argument in Proposition E.4 of \citet{fish_2025_stable}. Trial 1 achieves a substantial improvement, with a creative technique not applied anywhere in \citet{fish_2025_stable}. (Briefly, the LLM considers a dominating set of a tournament on the $g$ goods, and uses it to construct a $(t,u)$-stable menu. For the full proof see the data release in \cref{app:code_data_release}.) 

Regarding time taken, the LLM takes substantially less time to prove upper bounds than lower bounds (43m35s less on average, $p < 0.001$). This effect appears to be driven by the LLM writing more code (e.g.~invoking MILP solvers) for \textsc{Lower} than \textsc{Upper}. Unlike in the lower bound case, there is no significant difference in time taken by the \textsc{NoContext} and \textsc{WithContext} prompts. Finally, the best upper bound also took the most time to produce (31m 26s).

\paragraph{Conclusion.} It is, of course, difficult to isolate causes of particular LLM behaviors, especially in settings like this one in which the LLM's full CoT is not even accessible (let alone the model internals). Still, we conclude with some informal observations consistent with the above data:
\begin{enumerate}[label=(\arabic*)]
  \item For problems for which the LLM can make progress by pushing known techniques (here: lower bounds), providing human intuition does not appear to be necessary: the LLM can identify such promising directions on its own. 
  \item\label{point:lazy} The LLM is ``lazy'': it is ``satisfied'' with any progress, even if it is clearly trivial. Providing human intuition may counteract this effect by encouraging the LLM to be more ``ambitious'' / have better ``taste'' (here: upper bounds). 
\end{enumerate}

\section{Does automated multi-turn interaction help?}\label{sec:autoresearcher}

So far, we have demonstrated that a single-shot prompt is sufficient for eliciting improvements on \citet{fish_2025_stable} from the LLM. However, one limitation of this approach is the LLM's ``laziness'' (see \cref{point:lazy} above). For example, for the two lower bound trials that yielded $\gtrapprox 7/3 \approx 2.3$, it seems likely that a simple follow-up message such as \emph{``Can you improve this further?''} might have yielded the $\gtrapprox 2.5$ bounds found in other trials.

In this section, we experiment with a simple multi-turn ``autoresearcher'' pipeline, in which a second LLM agent ``supervises'' the main LLM research agent. 

\subsection{Pipeline architecture}

\input{pipeline_figure.tex}

See \cref{fig:pipeline} for an illustration. The pipeline consists of three agents, described in more detail below. For the full prompts see \cref{prompts_autoresearcher}. 

The first agent, Researcher, is given a prompt stating a research question, and (after an extended thinking period) outputs its result as freetext. It operates via a single GPT-5.5 Pro query, using the OpenAI API (as opposed to the ChatGPT web interface). We set the reasoning effort to \texttt{xhigh} and equip the LLM with two tools: OpenAI's built-in web-browsing tool, and a custom \texttt{run\_python} tool.\footnote{OpenAI does provide built-in tools for Python code execution---one Python code interpreter and one general shell environment. However, we encountered technical issues with both options, similar to what is described in this bug report: \url{https://community.openai.com/t/container-is-expired-error-in-openai-responses-api-stream/1321773}.} The \texttt{run\_python} tool gives the LLM access to a Docker container with 2GB memory, 120s timeout, and Python 3.11 with standard packages (numpy, scipy, z3-solver, pulp, ortools). Typical Researcher queries use 1 million input tokens, 20{,}000 reasoning tokens, and 50{,}000 output tokens.

The second agent, Extractor, is given the Researcher's output and asked to return the bound obtained by the Researcher (if applicable). It is implemented via a single GPT-5.5 API call. For the lower bounds, the Extractor is additionally asked to extract from the Researcher's response a specific instantiation of the construction (if applicable), which is fed into a verifier script. The verifier script takes as input a structured representation of a menu selection problem, alongside parameters $t,u \in \mathbb{N}$, and outputs whether a $(t,u)$-stable menu exists. The Extractor and Verifier are not particularly load-bearing; we include them in the spirit of prototyping (e.g. more complex workflows could involve verifying proofs with Lean). 

The third agent, Supervisor, is tasked with writing prompts to ``encourage'' the Researcher to make progress. It is implemented via a single Claude Opus 4.7 call.\footnote{Claude Opus 4.7 was selected for its strong agentic and prompt-writing capabilities. Though this model is less strong at mathematics research, it is sufficiently capable to understand the content.} The Supervisor is given: (1) information about the research question (2) the Researcher's prompt, summarized CoT, and output, (3) the Extractor's output, (4) memory notes written by the Supervisor in the previous turn. Then, the Supervisor's task is to output a new prompt for the Researcher. 

To illustrate, below is an excerpt from the Supervisor's (lower bound) prompt:
\begin{prompt}
When the subagent shows a promising pattern, give it concrete next steps (perturb cohort sizes, try a different underlying symmetry, recheck a specific menu cardinality). When it is stuck or keeps regenerating the known g = 7 / 23/11 construction, pivot -- change g, the preference structure, or the combinatorial template. After a new ratio is verified, do not stop: record it as the new bar and direct the next attempt at exceeding it.
\end{prompt}

And below is an excerpt from an example Supervisor-written prompt for the Researcher:
\begin{prompt}
A previous subagent proposed [...] This does not count as progress for our goal. [...] # Concrete directions to try [...] Direction 1 (recommended): Try g = 8 with a Z/8 cyclic structure. More goods give more potential orbit types. [...]
\end{prompt}

\subsection{Results}

We conduct three rollouts of the autoresearcher pipeline, testing each of the three prompts as the initial Researcher prompt: \textsc{NoContextLower}, \textsc{WithContextLower}, and \textsc{WithContextUpper}. Each rollout consists of 10 Researcher turns. 

\input{rollout_lower_with.tex}

\paragraph{Positive results.} \cref{tab:rollout_lower_with} displays the results for \textsc{WithContextLower}. As intended, the Supervisor successfully ``encourages'' the Researcher to seek out successively better lower bounds. The bounds obtained in turns 2--5 outperform \citet{fish_2025_stable}, and the bound from turn 5 ($\frac{u-1}{t-1} \gtrapprox 8/3 \approx 2.67$) outperforms the best bound from the single-shot pipeline ($\frac{u-1}{t-1} \gtrapprox 5/2 = 2.5$, see \cref{tab:llm_lower_bounds}). 

Beyond providing improved lower bounds, the LLM's $\frac{u-1}{t-1} \gtrapprox 8/3$ result also resolves a minor open question from \citet{fish_2025_stable}. Recall from \cref{sec:human_intuition} that \citet{fish_2025_stable}'s $\gtrapprox 23/11$ construction, as well as the LLM's $\gtrapprox 7/3$ and $\gtrapprox 5/2$ constructions, exhibit \emph{$(k,k+1)$-gaps}, for $k\!=\!2$ and $k\!=3\!$ respectively. Remark 5.6 of \citet{fish_2025_stable} asks whether all such constructions must exhibit \emph{$(k,k+1)$-gaps}. The LLM's $\gtrapprox 8/3$ construction does not satisfy this property, thereby resolving this open question via counterexample. 

\paragraph{Negative results.} In the \textsc{NoContextLower} and \textsc{WithContextUpper} experiments, the autoresearch pipeline gets ``stuck''. For \textsc{NoContextLower}, the Researcher does not beat the bounds from \citet{fish_2025_stable}, and for \textsc{WithContextUpper}, the Researcher at best extends $\frac{u-1}{t-1} \ge g-2$ to incomplete preferences (a near-trivial improvement, see \cref{sec:human_intuition}). With only one rollout per prompt type, it is unclear whether these results can be attributed to the prompt, or are simply noise. Still, we report some potential failure modes:
\begin{enumerate}[label=(\arabic*)]
  \item For \textsc{NoContextLower}, the Supervisor repeatedly encourages the Researcher to computationally search for constructions with cyclic structure, however, the pair fixates on using SMT/SAT solvers, rather than MILP/LP solvers (which appear to be more efficient in this context). 
  \item For \textsc{NoContextLower}, the (less mathematically capable) Supervisor appears to inject its own overly-specific hypotheses about promising constructions (e.g., asking for constructions based on $AG(2,3)$ or the Petersen graph). This does not happen in \textsc{WithContextLower}, where the Supervisor is a little more ``high-level'' (e.g., just asking for higher $g$ values). 
  \item For both \textsc{NoContextLower} and \textsc{WithContextUpper}, in later turns, the Supervisor appears to anchor on the progress rate of earlier turns. For example, the Supervisor's next steps in \textsc{WithContextLower} are more ``ambitious'' than those in \textsc{NoContextLower}. 
\end{enumerate}

We reiterate that these effects could be driven by noise, rather than by the prompts or harness. Still, these results suggest that Supervisor-type agents may be more effective when nudged to be more ambitious and high-level in their feedback. 

\section{Does the LLM outperform a first-year PhD student?}\label{sec:first-year}

\input{from_original_bounds.tex}

AI is changing the way we do research. We can ask it questions like we would a colleague, and we can delegate tasks to it like we would a coauthor. These dynamics may alter how we collaborate. To better understand such effects, in this section, we measure the counterfactual impact of adding an LLM to a collaboration, as opposed to a first-year PhD student, again using \cite{fish_2025_stable} as a testbed. 

To do this, we leverage the specific way that the collaboration in \citet{fish_2025_stable} unfolded. The project began as a collaboration between the senior authors G and H, which paused in 2020. They had written a five-page unpublished manuscript, GH'20, summarizing their progress.\footnote{GH'20 proved that $\frac{u-1}{t-1} \ge 2$ was a necessary condition (when $g \ge 3$), and that $\frac{u-1}{t-1} \ge g-1$ was a sufficient condition. Additionally, for $g = 4$, GH'20 obtained the upper bound $\frac{u-1}{t-1} \ge g-2 =2$, giving a complete characterization. Finally, GH'20 had conjectured that $\frac{u-1}{t-1} \ge 2$ was sufficient, which \citealt{fish_2025_stable} later disproved with the $\frac{u-1}{t-1} \gtrapprox 23/11$ bound for $g \ge 7$.} F (a first-year PhD student) joined the collaboration after this point, eventually resulting in \cite{fish_2025_stable}. Thus, using GH'20, we can compare two workflows: first, the baseline workflow of collaborating with a first-year PhD student to improve on GH'20 (resulting in \cite{fish_2025_stable}), and second, the workflow of asking an LLM to improve on GH'20.\footnote{Here, we focus on the single-shot setting, which yields a lower bound of the LLM's usefulness. A proper interactive ``collaboration'' with the LLM is likely more realistic, but also difficult to test experimentally.} 

One potential concern is that the LLM might ``cheat'' at the task of improving on GH'20 by drawing a connection to \citet{fish_2025_stable}. Fortunately, GH'20 used sufficiently different language and notation to describe the problem so that this seems unlikely (but of course, not possible to rule out). Indeed, across all experiments from this section, none of the reasoning summaries or LLM outputs explicitly reference \cite{fish_2025_stable}, or seek it out.\footnote{Example excerpt: \emph{``I don't need any current information or web results. The focus is on solving the math problem based on the uploaded paper.''}.}

We collect three samples each of \textsc{NoContextLower} and \textsc{NoContextUpper}, with the arXiv URL replaced by an attachment of GH'20. As in \cref{sec:human_intuition}, we use ChatGPT-5.5 Pro with Extended Thinking via the web interface, with browsing enabled, and personalization and memory disabled. The results of the six trials are displayed in \cref{tab:llm_from_original_lower,tab:llm_from_original_upper}. 

\paragraph{Upper bounds.} For the upper bounds, in all three trials, the LLM matches the $\frac{u-1}{t-1} \ge g-2$ bound from \citet{fish_2025_stable}. (For the comparisons here, we assume complete preferences, because the problem statement in GH'20, while otherwise identical, requires complete preferences and nonempty menus.) That is, the LLM is as effective as the first-year PhD student.

\paragraph{Lower bounds.} For the lower bounds, the LLM does not make progress beyond the trivial bound presented in GH'20. In particular, it does not beat the $\frac{u-1}{t-1} \gtrapprox 23/11$ lower bound from \citealt{fish_2025_stable}. This may be due to the LLM ``anchoring'' on GH'20's false conjecture that $\frac{u-1}{t-1} \ge 2$ ought to be a sufficient condition, in which case no lower bound improvements would be possible.

To mitigate this effect, we run three additional trials with the following sentence appended to the prompt (Conjecture 3 in GH'20 is the aforementioned false conjecture):
\begin{prompt}
  (In particular, ignore Conjecture 3, which is likely false.) 
\end{prompt}

The results are displayed in \cref{tab:llm_from_original_without_conjecture_lower}. In two out of the three trials, the LLM continues to fail to make progress. The reasoning summaries reveal attempts to use MILP/LP solvers, but the LLM does not push these techniques far enough to obtain an improved bound. In the third trial, the LLM claims an improved bound of $\frac{u-1}{t-1} \gtrapprox \frac{\log{g}}{\log{\log{g}}}$ (for sufficiently large $g$) using the probabilistic method, however, the proof does not appear to be correct.

Overall, these results indicate that the LLM is slightly less effective than the first-year PhD student. That said, the PhD student's advantage appears to be fragile; the lower bound reasoning summaries reveal that the LLM had the ``idea'' to search for constructions computationally, and \cref{sec:human_intuition,sec:autoresearcher} reveal that the LLM is adept at producing viable constructions when it exerts sufficient ``effort''. 

\printbibliography

\appendix

\crefalias{section}{appendix}
\crefname{appendix}{Appendix}{Appendices}
\Crefname{appendix}{Appendix}{Appendices}

\section{Code and data}\label{app:code_data_release}

The code is publicly available here: \url{https://github.com/sara-fish/stable_menus_ai_enabled_progress}. The LLM-written results can be viewed in this dashboard: \url{https://sara-fish.github.io/stable_menus_ai_enabled_progress/}.

\paragraph{Data collection.} For \cref{sec:human_intuition}, the data was collected between May 24, 2026 and May 26, 2026. For \cref{sec:autoresearcher}, the data was collected between May 25, 2026 and May 27, 2026. For \cref{sec:first-year}, the data was collected on May 30, 2026 and May 31, 2026.

\input{prompts.tex}

\end{document}

%% file: baselines_figure.tex
\usetikzlibrary{decorations.pathmorphing}

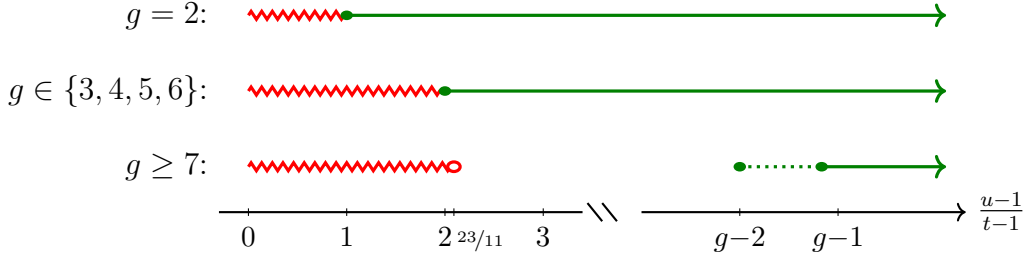
\begin{figure}[h]
\centering
\begin{tikzpicture}[xscale=1.3, yscale=1]
  \tikzset{zigzag red/.style={red, very thick, decorate, decoration={zigzag, segment length=4pt, amplitude=1.5pt}}}
  \def\xmin{-0.3}
  \def\breakL{3.4}
  \def\breakR{4.0}
  \def\gminus{5.0}
  \def\gone{6.0}
  \def\xmax{7.3}

  \draw[thick] (\xmin,0) -- (\breakL,0);
  \draw[->, thick] (\breakR,0) -- (\xmax,0) node[right] {$\frac{u-1}{t-1}$};

  \draw[thick] (\breakL+0.05,0.12) -- (\breakL+0.20,-0.12);
  \draw[thick] (\breakL+0.20,0.12) -- (\breakL+0.35,-0.12);

  \foreach \x in {0,1,2,3} {
    \draw (\x,0.05) -- (\x,-0.05) node[below, font=\small] {\x};
  }
  \draw (\gminus, 0.05) -- (\gminus, -0.05) node[below, font=\small] {$g{-}2$};
  \draw (\gone, 0.05) -- (\gone, -0.05) node[below, font=\small] {$g{-}1$};
  \draw (23/11, 0.05) -- (23/11, -0.05) node[below=1pt, xshift=10pt, font=\scriptsize] {$\nicefrac{23}{11}$};

  \def\ytwo{2.6}
  \node[left] at (\xmin, \ytwo) {$g = 2$:};
  \draw[zigzag red] (0, \ytwo) -- (1, \ytwo);
  \draw[green!50!black, very thick, ->] (1, \ytwo) -- (\xmax-0.2, \ytwo);
  \fill[green!50!black] (1, \ytwo) circle (1.8pt);

  \def\ymid{1.6}
  \node[left] at (\xmin, \ymid) {$g \in \{3,4,5,6\}$:};
  \draw[zigzag red] (0, \ymid) -- (2, \ymid);
  \draw[green!50!black, very thick, ->] (2, \ymid) -- (\xmax-0.2, \ymid);
  \fill[green!50!black] (2, \ymid) circle (1.8pt);

  \def\yseven{0.6}
  \node[left] at (\xmin, \yseven) {$g \ge 7$:};
  \draw[zigzag red] (0, \yseven) -- (23/11, \yseven);
  \draw[red, very thick, fill=white] (23/11, \yseven) circle (1.8pt);
  \draw[green!50!black, very thick, dotted] (\gminus, \yseven) -- (\gone-1/6, \yseven);
  \fill[green!50!black] (\gminus, \yseven) circle (1.8pt);
  \draw[green!50!black, very thick, ->] (\gone-1/6, \yseven) -- (\xmax-0.2, \yseven);
  \fill[green!50!black] (\gone-1/6, \yseven) circle (1.8pt);

\end{tikzpicture}
\caption{Visualization of the existence bounds obtained in \citet{fish_2025_stable}. Red zigzag lines indicate values of $\frac{u-1}{t-1}$ for which $(t,u)$-stable menus are not guaranteed to exist. Green solid lines indicate values of $\frac{u-1}{t-1}$ for which $(t,u)$-stable menus exist for all menu selection problems. Green dotted lines indicate that existence is only guaranteed for menu selection problems with complete preferences.}
\label{fig:bounds}
\end{figure}

%% file: lower_bounds.tex
\begin{table}[p]
\centering
\small
\renewcommand{\arraystretch}{1.3}
\begin{tabular}{lclll}
\hline
Treatment & Trial & Time taken & Bound found & Regime \\
\hline
\textsc{NoContextLower}   & 1 & 86m 4s  & $\frac{u-1}{t-1} \gtrapprox \bfrac{5}{2} = 2.5$ & $g \ge 9$ \\
\textsc{NoContextLower}   & 2 & 72m 21s & $\frac{u-1}{t-1} \gtrapprox \bfrac{5}{2} = 2.5$ & $g \ge 9$ \\
\textsc{NoContextLower}   & 3 & 84m 17s & $\frac{u-1}{t-1} \gtrapprox \bfrac{5}{2} = 2.5$ & $g \ge 9$ \\
\textsc{NoContextLower}   & 4 & 84m 8s  & $\frac{u-1}{t-1} \gtrapprox \bfrac{7}{3} \approx 2.3$ & $g \ge 9$ \\
\textsc{NoContextLower}   & 5 & 61m 8s  & $\frac{u-1}{t-1} \gtrapprox \bfrac{5}{2} = 2.5$ & $g \ge 9$ \\
\hline
\textsc{WithContextLower} & 1 & 57m 53s & $\frac{u-1}{t-1} \gtrapprox \bfrac{5}{2} = 2.5$ & $g \ge 9$ \\
\textsc{WithContextLower} & 2 & 42m 59s & $\frac{u-1}{t-1} \gtrapprox \bfrac{5}{2} = 2.5$ & $g \ge 9$ \\
\textsc{WithContextLower} & 3 & 64m 8s  & $\frac{u-1}{t-1} \gtrapprox \bfrac{5}{2} = 2.5$ & $g \ge 9$ \\
\textsc{WithContextLower} & 4 & 64m 11s & $\frac{u-1}{t-1} \gtrapprox \bfrac{7}{3} \approx 2.3$ & $g \ge 9$ \\
\textsc{WithContextLower} & 5 & 67m 52s & $\frac{u-1}{t-1} \gtrapprox \bfrac{5}{2} = 2.5$ & $g \ge 9$ \\
\hline
\hline
\citet[Theorem 1.4]{fish_2025_stable} & --- & --- & $\frac{u-1}{t-1} \gtrapprox \bfrac{23}{11} \approx 2.1$ & $g \ge 7$ \\
\hline
\end{tabular}
\caption{Lower bounds found and time taken by the LLM for each of the ten trials (five \textsc{NoContextLower}, five \textsc{WithContextLower}). The last row displays the $g \ge 7$ lower bound from \citet{fish_2025_stable}. Across all ten trials, the LLM improves on the bound from the paper (when $g \ge 9$). }
\label{tab:llm_lower_bounds}
\end{table}

%% file: upper_bounds.tex
\begin{table}[p]
\centering
\small
\renewcommand{\arraystretch}{1.3}
\begin{tabular}{lcllc}
\hline
Treatment & Trial & Time taken & Bound found & Preferences \\
\hline
\textsc{NoContextUpper}   & 1 & 21m 0s  & $\frac{u-1}{t-1} \ge g-2$                       & incomplete \\
\textsc{NoContextUpper}   & 2 & 19m 58s & $\frac{u-1}{t-1} \ge g-2$                       & incomplete \\
\textsc{NoContextUpper}   & 3 & 20m 34s & $\frac{u-1}{t-1} \gtrapprox g-2 + \frac{5}{g+1}$       & incomplete \\
\textsc{NoContextUpper}   & 4 & 25m 39s & $\frac{u-1}{t-1} \ge g-2$                       & incomplete \\
\textsc{NoContextUpper}   & 5 & 26m 34s &  $\frac{u-1}{t-1} \gtrapprox g-2 + \frac{5}{g+1}$                      & incomplete \\
\hline
\rowcolor{gray!20}
\textsc{WithContextUpper} & 1 & 31m 26s & $\frac{u-1}{t-1} \gtrapprox \frac{g}{2} + \log_2(g+1)$ & incomplete \\
\textsc{WithContextUpper} & 2 & 22m 8s  & $\frac{u-1}{t-1} \gtrapprox g-2 + \frac{5}{g+1}$       & incomplete \\
\rowcolor{gray!20}\textsc{WithContextUpper} & 3 & 26m 8s  & $\frac{u-1}{t-1} \gtrapprox    g-2 + \frac{4}{g}$        & incomplete \\
\textsc{WithContextUpper} & 4 & 25m 30s & $\frac{u-1}{t-1} \gtrapprox    g-2 + \frac{5}{g+1}$      & incomplete \\
\textsc{WithContextUpper} & 5 & 30m 12s & $\frac{u-1}{t-1} \ge g-2 $       & incomplete \\
\hline
\hline
\citet[Theorem 1.5]{fish_2025_stable} & --- & --- & $\frac{u-1}{t-1} \ge g-2$                       & complete only \\
\citet[Proposition E.4]{fish_2025_stable} & --- & --- & $\frac{u-1}{t-1} \gtrapprox g-\frac{7}{6}$ & incomplete \\
\hline
\end{tabular}
\caption{Upper bounds found and time taken by the LLM for each of the ten trials (five \textsc{NoContextUpper}, five \textsc{WithContextUpper}). The last two rows display two upper bound results from \citet{fish_2025_stable}. The LLM makes trivial improvements in eight out of the ten trials; the remaining two nontrivial improvements are highlighted. All bounds hold for $g \ge 7$.}
\label{tab:llm_upper_bounds}
\end{table}

%% file: pipeline_figure.tex

\begin{figure}[t]
  \centering
  \begin{tikzpicture}[
      node distance=12mm and 16mm,
      font=\small,
      stage/.style={
        rectangle, rounded corners, draw=black!55, line width=0.8pt,
        fill=black!8, text width=30mm, align=center,
        minimum height=15mm, inner sep=4pt,
      },
      tool/.style={
        rectangle, rounded corners, draw=black!45, line width=0.7pt,
        fill=black!4, align=center, font=\scriptsize,
        inner sep=3pt, minimum height=6mm,
      },
      flow/.style={-{Stealth[length=2.4mm]}, line width=0.9pt},
      toolflow/.style={{Stealth[length=1.8mm]}-{Stealth[length=1.8mm]},
        line width=0.7pt, draw=black!55},
    ]

    \node[stage] (research) {%
      \textbf{Researcher}\\[2pt]
      {\scriptsize\color{black!60}GPT-5.5 Pro}%
    };
    \node[tool, above left=4mm and 11mm of research.west, anchor=east] (websearch)
      {web search};
    \node[tool, below left=4mm and 11mm of research.west, anchor=east] (runpython)
      {run python};

    \draw[toolflow] (websearch.east) -- (research);
    \draw[toolflow] (runpython.east) -- (research);

    \node[stage, right=of research] (verify) {%
      \textbf{Extractor}\\[2pt]
      {\scriptsize\color{black!60}GPT-5.5}%
    };
    \node[stage, right=of verify] (super) {%
      \textbf{Supervisor}\\[2pt]
      {\scriptsize\color{black!60}Claude Opus 4.7}%
    };

    \draw[flow] (research) -- (verify);
    \draw[flow] (verify) -- (super);

    \draw[flow] (super.south) -- ++(0,-13mm) -| (research.south);

  \end{tikzpicture}
  \caption{Illustration of the ``autoresearcher'' pipeline.}
  \label{fig:pipeline}
\end{figure}
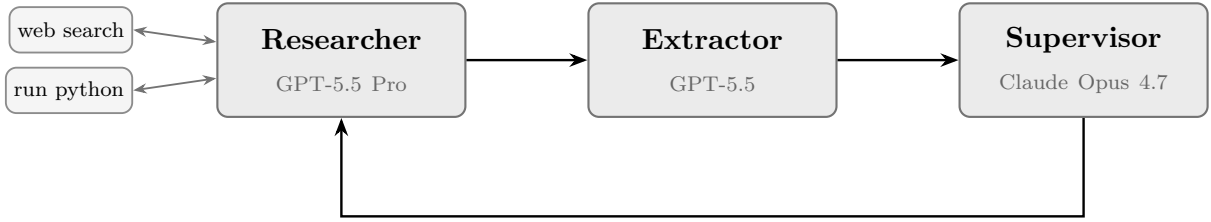

%% file: rollout_lower_with.tex
\begin{table}[t]
\centering
\small
\renewcommand{\arraystretch}{1.3}
\begin{tabular}{clll}
\hline
Turn & Time taken & Bound found & $(g,t,u,n)$ \\
\hline
1 & 102m 22s & $\frac{u-1}{t-1} \gtrapprox \bfrac{23}{11} \approx 2.09$ & $(7,12,23,70)$ \\
2 & 82m 35s & $\frac{u-1}{t-1} \gtrapprox \bfrac{13}{6} \approx 2.17$ & $(9,19,39,54)$ \\
3 & 38m 31s & $\frac{u-1}{t-1} \gtrapprox \bfrac{7}{3} \approx 2.33$ & $(9,10,21,27)$ \\
4 & 38m 53s & $\frac{u-1}{t-1} \gtrapprox \bfrac{5}{2} \approx 2.50$ & $(9,9,20,18)$ \\
\rowcolor{gray!20}5 & 104m 7s & $\frac{u-1}{t-1} \gtrapprox \bfrac{8}{3} \approx 2.67$ & $(9,4,8,27)$ \\
6 & 175m 25s & $\frac{u-1}{t-1} \gtrapprox \bfrac{8}{3} \approx 2.67$ & $(9,4,8,27)$ \\
7 & 140m 12s & $\frac{u-1}{t-1} \gtrapprox \bfrac{8}{3} \approx 2.67$ &$(9,4,8,27)$ \\
8 & 155m 16s & none found & --- \\
9 & 146m 56s & none found & --- \\
10 & 166m 30s & none found & --- \\
\hline
\hline
\citet[Theorem 1.4]{fish_2025_stable} & --- & $\frac{u-1}{t-1} \gtrapprox \bfrac{23}{11} \approx 2.09$ & $(7,12,23,70)$ \\
\hline
\end{tabular}
\caption{Lower bounds found (with construction parameters $g,t,u,n$) and time taken by the autoresearcher, with initial prompt \textsc{WithContextLower}, for ten sequential turns. The last row displays the $g \ge 7$ lower bound from \citet{fish_2025_stable}. The LLM improves on the results from the one-shot queries from \cref{tab:llm_lower_bounds} in turn 5 (highlighted).}
\label{tab:rollout_lower_with}
\end{table}

%% file: from_original_bounds.tex
\begin{table}[p]
\centering
\renewcommand{\arraystretch}{1.3}
\begin{tabular}{lll}
\hline
Trial & Time taken & Bound found \\
\hline
1 & 58m 11s & $\frac{u-1}{t-1} \ge 2$ \\
2 & 62m 57s & $\frac{u-1}{t-1} \ge 2$ \\
3 & 79m 54s & $\frac{u-1}{t-1} \ge 2$ \\
\hline
\hline
GH'20                                 & --- & $\frac{u-1}{t-1} \ge 2$ \\
\citet[Theorem 1.4]{fish_2025_stable} & --- & $\frac{u-1}{t-1} \gtrapprox \bfrac{23}{11} \approx 2.1$ \\
LLM, one-shot (best)      & 42m59s   & $\frac{u-1}{t-1} \gtrapprox \bfrac{5}{2} = 2.5$ \\
LLM, autoresearcher & 366m28s  & $\frac{u-1}{t-1} \gtrapprox \bfrac{8}{3} \approx 2.67$ \\
\hline
\end{tabular}
\caption{Lower bounds found and time taken by the LLM for each of the three trials of \textsc{NoContextLower}, with GH'20 provided instead of \cite{fish_2025_stable}. The final four rows display baselines (see \cref{tab:llm_lower_bounds,tab:rollout_lower_with}).}
\label{tab:llm_from_original_lower}
\end{table}

\begin{table}[p]
\centering
\renewcommand{\arraystretch}{1.3}
\begin{tabular}{lll}
\hline
Trial & Time taken & Bound found \\
\hline
1 & 29m 14s & $\frac{u-1}{t-1} \ge g-2$ \\
2 & 34m 10s & $\frac{u-1}{t-1} \ge g-2$ \\
3 & 27m 19s & $\frac{u-1}{t-1} \ge g-2$ \\
\hline
\hline
GH'20                                 & --- & $\frac{u-1}{t-1} \ge g-1$ \\
\citet[Theorem 1.5]{fish_2025_stable} & --- & $\frac{u-1}{t-1} \ge g-2$  \\
LLM, one-shot (best)      & 31m26s & $\frac{u-1}{t-1} \gtrapprox \frac{g}{2} + \log_2(g+1)$ \\
LLM, autoresearcher & 916m21s & $\frac{u-1}{t-1} \ge g-2$ \\
\hline
\end{tabular}
\caption{Upper bounds found and time taken by the LLM for each of the three trials of \textsc{NoContextUpper}, with GH'20 provided instead of \cite{fish_2025_stable}. The final four rows display baselines (see \cref{tab:llm_upper_bounds}).}
\label{tab:llm_from_original_upper}
\end{table}

\begin{table}[p]
\centering
\renewcommand{\arraystretch}{1.3}
\begin{tabular}{lll}
\hline
Trial & Time taken & Bound found \\
\hline
1 & 90m 10s & $\frac{u-1}{t-1} \ge 2$ \\
2 & 85m 33s & $\frac{u-1}{t-1} \ge 2$ \\
3 & 88m 39s & invalid (flawed proof claiming $\frac{u-1}{t-1} \gtrapprox \frac{\log{g}}{\log{\log{g}}}$ for suff.~large $g$) \\
\hline
\end{tabular}
\caption{Lower bounds found and time taken by the LLM for each of the three trials of \textsc{NoContextLower}, with GH'20 provided instead of \cite{fish_2025_stable}, and with an additional instruction to ignore a false conjecture from GH'20. For baselines see \cref{tab:llm_from_original_lower}.}
\label{tab:llm_from_original_without_conjecture_lower}
\end{table}

%% file: prompts.tex
\section{Prompts}\label{prompts}

\subsection{Single-turn prompts}\label{prompts_single_turn}

These prompts were entirely human-written. 

\subsubsection{{NoContextLower}}

\begin{prompt}
https://arxiv.org/abs/2402.11370

I would like you to work on deriving improvements for the lower bound constructions, that is, the necessary condition bounds. Do not work on upper bound proofs, because another agent has that covered. Keep working until you have an actual lower bound construction better than the bounds in the paper. No need to report back with partial progress not yet constituting an improved lower bound. All of the lower bounds in the paper in fact are very loose, so progress is tractable.
\end{prompt}

\subsubsection{{NoContextUpper}}

\begin{prompt}
https://arxiv.org/abs/2402.11370

I would like you to work on deriving improvements for the upper bounds, that is, the sufficient condition bounds. Do not work on lower bound constructions, because another agent has that covered. Keep working until you have an actual upper bound proof better than the bounds in the paper. No need to report back with partial progress not yet constituting a proof. All of the upper bounds in the paper in fact are very loose, so progress is tractable.
\end{prompt}

\subsubsection{{WithContextLower}}

\begin{prompt}
https://arxiv.org/abs/2402.11370

This paper presents a variety of bounds that aim to understand necessary and sufficient conditions for the existence of stable menus. When the number of goods g satisfies g <= 6, the paper provides a complete characterization. For g= 2, it shows that (t,u)-stable menus exist for all menu selection problems if and only if u >= t. For g = 3,4,5,6, it shows that (t,u)-stable menus exist for all menu selection problems if and only if u >= 2t-1. For g >= 7, the lower and upper bounds do not match. The paper shows using a lower bound construction that it is necessary for u >= 23 \lfloor (t-1) / 11 \rfloor to have (t,u)-stable menus exist for all menu selection problems. The paper also proves, for all g >= 7, that (u-1)/(t-1) >= g-2 is a sufficient condition (with complete preferences).

I would like you to work on deriving improvements for the lower bound constructions, that is, the necessary condition bounds. Do not work on upper bound proofs, because another agent has that covered. Because the paper already provides a complete characterization for g <= 6, progress for lower bounds involves looking at g >= 7. A good place to start would be to look at the lower bound constructions in the paper. They include the construction consisting of cycles 1>2>3, 2>3>1, 3>1>2 that show that u >= 2t-1 is a necessary condition for the existence of stable menus for g >= 3, and then the more complex construction in Section 5 showing that u >= 23 \lfloor (t-1)/11 \rfloor + 1 is necessary for the existence of stable menus for g >= 7. You'll notice in both cases, the constructions have a specific cyclic structure. One promising direction could be to continue to look for constructions of this form. Of course, you are welcome to pursue other techniques. Keep working until you have an actual lower bound construction better than the bounds in the paper. No need to report back with partial progress not yet constituting an improved lower bound. All of the lower bounds in the paper in fact are very loose, so progress is tractable.
\end{prompt}

\subsubsection{{WithContextUpper}}

\begin{prompt}
https://arxiv.org/abs/2402.11370

This paper presents a variety of bounds that aim to understand necessary and sufficient conditions for the existence of stable menus. When the number of goods g satisfies g <= 6, the paper provides a complete characterization. For g= 2, it shows that (t,u)-stable menus exist for all menu selection problems if and only if u >= t. For g = 3,4,5,6, it shows that (t,u)-stable menus exist for all menu selection problems if and only if u >= 2t-1. For g >= 7, the lower and upper bounds do not match. The paper shows using a lower bound construction that it is necessary for u >= 23 \lfloor (t-1) / 11 \rfloor to have (t,u)-stable menus exist for all menu selection problems. The paper also proves, for all g >= 7, that (u-1)/(t-1) >= g-2 is a sufficient condition (with complete preferences).

I would like you to work on deriving improvements for the upper bounds, that is, the sufficient condition bounds. Do not work on lower bound constructions, because another agent has that covered. Because the paper already provides a complete characterization for g <= 6, progress for upper bounds involves looking at g >= 7. A good place to start would be to look at the upper bound proofs in the paper. They include the aforementioned (u-1)/(t-1) = g-2 bound, but also the simple u >= (g-1)(t-1) argument in Appendix E, as well as the more refined u >= (g-1-eps)(t-1), also in Appendix E. These arguments arrive at sufficient conditions for the existence of (t,u)-stable menus for certain g by uncovering some sort of mathematical structure about the problem. The (g-2) bound uncovers one kind of structure (this "no gaps" idea), and the (g-1-eps) bound uncovers a different kind of structure (that, when small (t,u)-stable menus fail to exist, for u large, this implies that the menu selection problem has a specific cyclic structure). One promising direction could be to push either of these observations, or both simultaneously. Or, you are welcome to pursue other techniques. Keep working until you have an actual upper bound proof better than the bounds in the paper. No need to report back with partial progress not yet constituting a proof. All of the upper bounds in the paper in fact are very loose, so progress is tractable.
\end{prompt}

\subsection{Autoresearcher prompts}\label{prompts_autoresearcher}

These prompts were written by asking Claude Opus 4.7 to flesh out a human-written outline. To avoid AI detector false positives, the prompts can be found here: \url{https://github.com/sara-fish/stable_menus_ai_enabled_progress/blob/e12e54863e397c4b23e0cff1c0e7d21468d60855/autoresearcher/prompts.py}

%% file: refs.bib
@misc{feng_towards_2026,
	title = {Towards {Autonomous} {Mathematics} {Research}},
	url = {https://arxiv.org/abs/2602.10177v3},
	language = {en},
	urldate = {2026-06-06},
	journal = {arXiv.org},
	author = {Feng, Tony and Trinh, Trieu H. and Bingham, Garrett and Hwang, Dawsen and Chervonyi, Yuri and Jung, Junehyuk and Lee, Joonkyung and Pagano, Carlo and Kim, Sang-hyun and Pasqualotto, Federico and Gukov, Sergei and Lee, Jonathan N. and Kim, Junsu and Hou, Kaiying and Ghiasi, Golnaz and Tay, Yi and Li, YaGuang and Kuang, Chenkai and Liu, Yuan and Lin, Hanzhao and Liu, Evan Zheran and Nayakanti, Nigamaa and Yang, Xiaomeng and Cheng, Heng-Tze and Hassabis, Demis and Kavukcuoglu, Koray and Le, Quoc V. and Luong, Thang},
	month = feb,
	year = {2026},
}

@misc{zheng2026aicomathematicianacceleratingmathematicians,
      title={AI co-mathematician: Accelerating mathematicians with agentic AI}, 
      author={Daniel Zheng and Ingrid von Glehn and Yori Zwols and Iuliya Beloshapka and Lars Buesing and Daniel M. Roy and Martin Wattenberg and Bogdan Georgiev and Tatiana Schmidt and Andrew Cowie and Fernanda Viegas and Dimitri Kanevsky and Vineet Kahlon and Hartmut Maennel and Sophia Alj and George Holland and Alex Davies and Pushmeet Kohli},
      year={2026},
      eprint={2605.06651},
      archivePrefix={arXiv},
      primaryClass={cs.AI},
      url={https://arxiv.org/abs/2605.06651}, 
}

@misc{wagner2021constructions,
	title = {Constructions in combinatorics via neural networks},
	url = {http://arxiv.org/abs/2104.14516},
	doi = {10.48550/arXiv.2104.14516},
	urldate = {2026-06-06},
	publisher = {arXiv},
	author = {Wagner, Adam Zsolt},
	month = apr,
	year = {2021},
	note = {arXiv:2104.14516 [math.CO]},
}

@misc{novikov2025alphaevolve,
	title = {{AlphaEvolve}: {A} coding agent for scientific and algorithmic discovery},
	shorttitle = {{AlphaEvolve}},
	url = {http://arxiv.org/abs/2506.13131},
	doi = {10.48550/arXiv.2506.13131},
	urldate = {2026-06-06},
	publisher = {arXiv},
	author = {Novikov, Alexander and Vũ, Ngân and Eisenberger, Marvin and Dupont, Emilien and Huang, Po-Sen and Wagner, Adam Zsolt and Shirobokov, Sergey and Kozlovskii, Borislav and Ruiz, Francisco J. R. and Mehrabian, Abbas and Kumar, M. Pawan and See, Abigail and Chaudhuri, Swarat and Holland, George and Davies, Alex and Nowozin, Sebastian and Kohli, Pushmeet and Balog, Matej},
	month = jun,
	year = {2025},
	note = {arXiv:2506.13131 [cs.AI]},
}

@misc{tsoukalas_advancing_2026,
	title = {Advancing {Mathematics} {Research} with {AI}-{Driven} {Formal} {Proof} {Search}},
	url = {http://arxiv.org/abs/2605.22763},
	doi = {10.48550/arXiv.2605.22763},
	urldate = {2026-06-06},
	publisher = {arXiv},
	author = {Tsoukalas, George and Kovsharov, Anton and Shirobokov, Sergey and Surina, Anja and Firsching, Moritz and Bérczi, Gergely and Ruiz, Francisco J. R. and Suggala, Arun and Wagner, Adam Zsolt and Wieser, Eric and Yu, Lei and Huang, Aja and Horváth, Miklós Z. and Ferrauiolo, Andrew and Michalewski, Henryk and Grosu, Codrut and Hubert, Thomas and Balog, Matej and Kohli, Pushmeet and Chaudhuri, Swarat},
	month = may,
	year = {2026},
	note = {arXiv:2605.22763 [cs.AI]
version: 1},
}

@misc{nagda_reinforced_2026,
	title = {Reinforced {Generation} of {Combinatorial} {Structures}: {Ramsey} {Numbers}},
	shorttitle = {Reinforced {Generation} of {Combinatorial} {Structures}},
	url = {http://arxiv.org/abs/2603.09172},
	doi = {10.48550/arXiv.2603.09172},
	urldate = {2026-06-06},
	publisher = {arXiv},
	author = {Nagda, Ansh and Raghavan, Prabhakar and Thakurta, Abhradeep},
	month = apr,
	year = {2026},
	note = {arXiv:2603.09172 [math.CO]},
}

@misc{charton_patternboost_2024,
	title = {{PatternBoost}: {Constructions} in {Mathematics} with a {Little} {Help} from {AI}},
	shorttitle = {{PatternBoost}},
	url = {http://arxiv.org/abs/2411.00566},
	doi = {10.48550/arXiv.2411.00566},
	urldate = {2026-06-06},
	publisher = {arXiv},
	author = {Charton, François and Ellenberg, Jordan S. and Wagner, Adam Zsolt and Williamson, Geordie},
	month = nov,
	year = {2024},
	note = {arXiv:2411.00566 [math.CO]},
}

@misc{georgiev_mathematical_2025,
	title = {Mathematical exploration and discovery at scale},
	url = {http://arxiv.org/abs/2511.02864},
	doi = {10.48550/arXiv.2511.02864},
	urldate = {2026-06-06},
	publisher = {arXiv},
	author = {Georgiev, Bogdan and Gómez-Serrano, Javier and Tao, Terence and Wagner, Adam Zsolt},
	month = dec,
	year = {2025},
	note = {arXiv:2511.02864 [cs.NE]},
}

@misc{breen_ax-prover_2026,
	title = {Ax-{Prover}: {A} {Deep} {Reasoning} {Agentic} {Framework} for {Theorem} {Proving} in {Mathematics} and {Quantum} {Physics}},
	shorttitle = {Ax-{Prover}},
	url = {http://arxiv.org/abs/2510.12787},
	doi = {10.48550/arXiv.2510.12787},
	urldate = {2026-06-06},
	publisher = {arXiv},
	author = {Breen, Benjamin and Tredici, Marco Del and McCarran, Jacob and Mijares, Javier Aspuru and Yin, Weichen Winston and Sulimany, Kfir and Taylor, Jacob M. and Koppens, Frank H. L. and Englund, Dirk},
	month = may,
	year = {2026},
	note = {arXiv:2510.12787 [cs.AI]},
}

@article{wang_scientific_2023,
	title = {Scientific discovery in the age of artificial intelligence},
	volume = {620},
	copyright = {2023 Springer Nature Limited},
	issn = {1476-4687},
	url = {https://www.nature.com/articles/s41586-023-06221-2},
	doi = {10.1038/s41586-023-06221-2},
	number = {7972},
	urldate = {2026-06-06},
	journal = {Nature},
	author = {Wang, Hanchen and Fu, Tianfan and Du, Yuanqi and Gao, Wenhao and Huang, Kexin and Liu, Ziming and Chandak, Payal and Liu, Shengchao and Van Katwyk, Peter and Deac, Andreea and Anandkumar, Anima and Bergen, Karianne and Gomes, Carla P. and Ho, Shirley and Kohli, Pushmeet and Lasenby, Joan and Leskovec, Jure and Liu, Tie-Yan and Manrai, Arjun and Marks, Debora and Ramsundar, Bharath and Song, Le and Sun, Jimeng and Tang, Jian and Veličković, Petar and Welling, Max and Zhang, Linfeng and Coley, Connor W. and Bengio, Yoshua and Zitnik, Marinka},
	month = aug,
	year = {2023},
	note = {Publisher: Nature Publishing Group},
	pages = {47--60},
}

@inproceedings{fish_2025_stable,
shorthand = {FGH'25},
author = {Fish, Sara and Gonczarowski, Yannai A. and Hart, Sergiu},
title = {Stable Menus of Public Goods: A Matching Problem},
year = {2025},
isbn = {9798400719431},
publisher = {Association for Computing Machinery},
address = {New York, NY, USA},
url = {https://doi.org/10.1145/3736252.3742543},
doi = {10.1145/3736252.3742543},
booktitle = {Proceedings of the 26th ACM Conference on Economics and Computation},
pages = {348},
numpages = {1},
location = {Stanford University, Stanford, CA, USA},
series = {EC '25}
}

@misc{openai_openai_2026,
	title = {An {OpenAI} model has disproved a central conjecture in discrete geometry},
	url = {https://openai.com/index/model-disproves-discrete-geometry-conjecture/},
	urldate = {2026-05-31},
	journal = {OpenAI},
	author = {{OpenAI}},
	month = may,
	year = {2026},
}

@misc{bloom_sum-product_2026,
	title = {The sum-product conjecture is false for real numbers},
	url = {http://arxiv.org/abs/2605.28781},
	doi = {10.48550/arXiv.2605.28781},
	urldate = {2026-05-31},
	publisher = {arXiv},
	author = {Bloom, Thomas F. and Sawin, Will and Schildkraut, Carl and Zhelezov, Dmitrii},
	month = may,
	year = {2026},
	note = {arXiv:2605.28781 [math.NT]
version: 1},
}
